# Ground-truth "resting-state" signal provides data-driven estimation and correction for scanner distortion of fMRI time-series dynamics

Rajat Kumar[1], Liang Tan[2], Alan Kriegstein[2], Andrew Lithen[1], Jonathan R. Polimeni[3,4], Helmut H. Strey[1,5], Lilianne R. Mujica-Parodi[1,3,5]

1. Department of Biomedical Engineering, Stony Brook University, Stony Brook, NY—USA.
2. ALA Scientific Instruments, Inc., Farmingdale, NY—USA.
3. Athinoula A. Martinos Center for Biomedical Imaging, Department of Radiology, Harvard Medical School, Massachusetts General Hospital, Charlestown, MA—USA.
4. Harvard-Massachusetts Institute of Technology Division of Health Sciences and Technology, Cambridge, MA—USA.
5. Laufer Center for Physical and Quantitative Biology, Stony Brook University, Stony Brook, NY—USA.

For correspondence, please contact:

Lilianne R. Mujica-Parodi, Ph.D.
Director, Laboratory for Computational Neurodiagnostics
Professor, Department of Biomedical Engineering
Stony Brook University School of Medicine
Stony Brook, NY 11794-5281
Email: Lilianne.Strey@stonybrook.edu

And

Helmut H. Strey, Ph.D.
Associate Professor, Department of Biomedical Engineering
Stony Brook University School of Medicine
Stony Brook, NY  11794-5281
Email:  Helmut.Strey@stonybrook.edu

Abstract

The fMRI community has made great strides in decoupling neuronal activity from other physiologically induced $T_2^*$ changes, using sensors that provide a ground-truth with respect to cardiac, respiratory, and head movement dynamics. However, blood oxygenation level-dependent (BOLD) time-series dynamics are also confounded by scanner artifacts, in complex ways that can vary not only between scanners but even, for the same scanner, between sessions. Unfortunately, the lack of an equivalent ground truth for BOLD time-series has thus far stymied the development of reliable methods for identification and removal of scanner-induced noise, a problem that we have previously shown to severely impact detection sensitivity of resting-state brain networks. To address this problem, we first designed and built a phantom capable of providing dynamic signals equivalent to that of the resting-state brain. Using the dynamic phantom, we then compared the ground-truth time-series with its measured fMRI data. Using these, we introduce data-quality metrics: Standardized Signal-to-Noise Ratio (ST-SNR) and Dynamic Fidelity that, unlike current measures such as temporal SNR (tSNR), can be directly compared across scanners. Dynamic phantom data acquired from four "best-case" scenarios: high-performance scanners with MR-physicist-optimized acquisition protocols, still showed scanner instability/multiplicative noise contributions of about 6–18% of the total noise. We further measured strong non-linearity in the fMRI response for all scanners, ranging between 8–19% of total voxels. To correct scanner distortion of fMRI time-series dynamics at a single-subject level, we trained a convolutional neural network (CNN) on paired sets of measured vs. ground-truth data. The CNN learned the unique features of each session's noise, providing a customized temporal filter. Tests on dynamic phantom time-series showed a 4- to 7-fold increase in ST-SNR and about 40–70% increase in Dynamic Fidelity after denoising, with CNN denoising outperforming both the temporal bandpass filtering and denoising using Marchenko-Pastur principal component analysis. Critically, we observed that the CNN temporal denoising pushes ST-SNR to a regime where signal power is higher than that of noise (ST-SNR > 1). Denoising human-data with ground-truth-trained CNN, in turn, showed markedly increased detection sensitivity of resting-state networks. These were visible even at the level of the single-subject, as required for clinical applications of fMRI.

# 1. Introduction

Large-scale investments in the identification of fMRI-derived biomarkers for brain-based disorders are a testament to the anticipated promise of fMRI as a neurodiagnostic tool. Yet even once clinical neuroscience establishes reliable biomarkers, a critical rate-limiting factor in the use of fMRI in clinical practice will be fMRI's poor signal/noise profile for single-subject level analyses. The task-free, "resting-state" paradigms most likely to be utilized in a clinical setting (because of their limited reliance on patient training, engagement, and compliance) only exacerbate this problem. Task-based designs, in principle, clearly delineate between activation in response to a task (signal) and activation during baseline (noise). However, task-free paradigms, by definition, lack the experimental manipulation that would typically be used to distinguish between *fluctuations of interest* (signal) from *fluctuations of nuisance* (noise) (DeDora et al. 2016). Without a principled way to distinguish between signal and noise, we lack the feedback necessary to optimize for one while removing the other, thereby limiting our ability to achieve the kind of advances in detection sensitivity required to enhance fMRI's utility in evaluating the single patient.

FMRI's signal is conventionally derived from the *blood oxygenation level-dependent* (BOLD) contrast. This activity represents regional time-varying changes in the concentration of deoxygenated hemoglobin, following neural-activity induced by exogenous stimuli or spontaneous fluctuations of the resting state. These time-varying changes reflect changes in apparent transverse relaxation time $T_2^*$, an MR parameter sensitive to levels of deoxyhemoglobin, and hence responsible for the observed BOLD contrast. Ideally, the value measured at each voxel at a given time point should only change in response to $T_2^*$ changes driven by neural activity (fluctuations of interest, signal). However, in practice, the measurement is dependent on a complex interaction between acquisition parameters (flip-angle: $\alpha$, echo-time: TE, repetition-time: TR), MR parameters (longitudinal relaxation time $T_1$, apparent transverse relaxation time $T_2^*$, proton density within a voxel) and background noise (Lauterbur 2000). Change in any of these parameters introduces variance (fluctuations of the nuisance, noise) in the observed voxel time-series. The difficulty of maintaining fidelity to actual (neuronal) time-series dynamics is made even more acute by the fact that BOLD contrast constitutes only a small fraction (typically, less than 5%) of the total measured signal.

Fluctuations of nuisance in the fMRI time-series originate from two sources: the individual being scanned (physiological noise, due primarily to cardiac, respiratory, and motion effects) as well as the scanner itself. Physiological processes like respiration or cardiac pulsations can cause changes in blood flow (affecting $T_1$ and $T_2^*$), and thus temporal variations in magnetization that might artifactually appear to be a BOLD effect. Subject head motion causes relative displacement of voxels leading to temporally correlated non-stationary noise and can induce spurious correlations in the resting-state analysis (Power et al. 2012). As significant as these artifacts are, the fact that cardiac, respiratory, and motion variables permit external measurements (e.g., ECG for heart rate) have permitted the field to develop an impressive array of well-validated methods with which to both identify and mitigate their influence. Examples of strategies for targeting physiological and motion confounds include: selecting acquisition parameters designed to permit thermal noise to dominate physiological noise (Wald and Polimeni 2017); techniques to address breathing-related field fluctuations both prospectively (Duerst et al. 2015) and at image reconstruction stage (Bollmann et al. 2017); use of simultaneously recorded measurement of heart-rate, respiration, and motion to retrospectively remove physiological confounds (Caballero-Gaudes and Reynolds 2017); and motion-correction implemented prospectively (Zaitsev et al. 2017) or retrospectively through registration.

In contrast, the lack of a ground truth for fMRI time-series has not permitted the same strategies for identification and removal of scanner-induced noise, which can vary not only between scanners of the same make and model, but even within the same scanner during different sessions. These fluctuations of nuisance originate from imperfections of the instrumentation and the electromagnetic fields used for the measurement and are normally referred to as "scanner instability." This nomenclature is, itself, potentially misleading, since detection-sensitivity of resting-state networks requires simultaneously amplifying fluctuations of interest while suppressing fluctuations of nuisance. Indeed, we have previously shown that typical methods that focus entirely on suppressing fluctuations (optimizing solely for scanner "stability"), such as temporal signal/noise (tSNR), actually *deoptimize* detection-sensitivity of resting-state networks, because the damped fluctuations include not only suppressed noise but also suppressed signal (DeDora et al. 2016).

Different approaches tackle the problem of minimizing scanner artifacts based upon models of MR-physics. Such methods include reducing the effects of eddy currents by the use of actively shielded gradients and pre-emphasis filters, the use of navigators and calibration echoes, or NMR probes(Kasper et al. 2015) that provide concurrent field monitoring with correction during image reconstruction. Yet modeling-based approaches, while valuable in their own right in terms of contributing to our understanding, can fall short as a practical tool for optimizing resting-state signal/noise (SNR). The reason for this is that they tend to oversimplify processes that, in an actual testing environment, are fundamentally complex—involving multiple factors, both known and unknown, which interact with one another in nonlinear and nonstationary ways. For example, scanner instabilities may be caused by variation in flip angle over time, imperfections in gradient system, heating, time-varying eddy current effects, or gain changes in transmit and receive chains (Greve et al. 2013; Liu 2016). Time-varying gradients in fast imaging methods, such as interleaved echo-planar imaging (EPI), require high-gradient amplitudes and slew-rates, pushing the scanner to its limits and causes image artifacts due to k-space trajectory deviations. Inhomogeneity in $B_0$ field and perturbations in gradient field cause eddy currents, ghosting, geometric distortions, errors in phase encoding leading to voxel displacement, gain-drifts, and other distortions (Jezzard and Clare 1999). While scanner instability is multiplicative, the impact of thermal/background noise on fMRI time-series is additive and can arise due to a random process like Brownian motion of ions in MR electronics or the human subject, external RF noise sources in the scanner room, or RF spikes dues to intermittent contact between metallic components (Greve et al. 2013; Liu 2016).

In a clinical setting involving decision-making for a single patient, the impact of errors that fluctuate over time and are signal dependent cannot be remedied by increasing sample size, under the assumption that signal amplifies while noise cancels. Longitudinal comparison of scans acquired pre and post treatment cannot be interpreted if both the subject and scanner are changing over time (for example, in using resting-state fMRI in pre-surgical localization, surgical planning in epilepsy, and identifying subjects with Alzheimer's Disease (Lee, Smyser, and Shimony 2013)). Moreover, biomarkers used at one site may be difficult to compare across other sites. Even in the research domain, recent years have seen a tremendous increase in efforts in pooling fMRI data for increasing sample size, enhancing statistical power for detecting subtle effects, including diverse populations and disease etiologies (Van Horn and Toga 2009), either via multi-site studies or data-sharing initiatives. Combining data from multiple sites presents an unavoidable challenge in the form of scanner-induced inter-site variability due to differences in field strength, imaging parameters, image reconstruction, or scanner manufacturer (Glover et al. 2012) and can lead to systematic confounds in time-series data. In one recent example (Friedman et al. 2008), between-site reliability showed median intra-class correlation of just *r*=0.22.

In summary, efforts to make the application of resting-state fMRI clinically useful must necessarily address SNR from the perspective of not only physiological, but scanner, artifact—and in ways that make sense given the ubiquity of task-free designs. While efforts to mitigate physiological artifact can and have benefited from external measurements (Caballero-Gaudes and Reynolds 2017), until recently such a strategy has not been available for scanner artifact. Static phantoms optimize purely for general stability (Friedman and Glover 2006), thereby suppressing the fluctuations responsible for resting-state signal. Moreover, the brain (non-static but, by definition, the unknown variable) likewise cannot serve as a calibration device. Finally, physics-based models cannot, in principle, approximate the impact of complex nonstationary distortion on time-series without empirical measurement of that distortion. To address these issues, we approached the problem from the perspective of creating a "brain-like" calibration device, capable of producing a dynamic ground-truth input signal similar to a typical resting-state time-series. Because such a device would provide a ground truth for both fluctuations of interest (signal) as well as fluctuations of nuisance (noise), it could permit optimization for signal-to-noise, rather than simply stability. Because of the consequent ability to obtain, and therefore compare, time-series distortion between true and measured time-series, we could develop a purely data-driven—rather than modeled—distortion correction. Doing so would potentially permit cleaning data of scanner-induced artifact while remaining agnostic with respect to the diversity of known and unknown sources of distortion and their behavior over time.

## 2. Results

**2.1.** *We designed and engineered a commercial-grade dynamic phantom capable of producing brain-like dynamic signals.* Our previous work (Rădulescu and Mujica-Parodi 2014; Mujica-Parodi, Cha, and Gao 2017) and those of others (Ciuciu et al. 2012) shows that healthy resting-state fMRI signals follow 1/f (pink noise) frequency spectra; therefore, our pseudo-brain "input" signal was engineered to achieve equivalent dynamics (custom dynamics can also be easily programmed). To create a dynamic signal, our phantom (**Fig. 1A**) uses difference in agarose gel concentration across voxels; the phantom, when rotated in-plane across a voxel during the data acquisition, produces a changing $T_2^*$ signal. Rotations occur at the start of each TR of a scan and are limited to around 250 milliseconds.

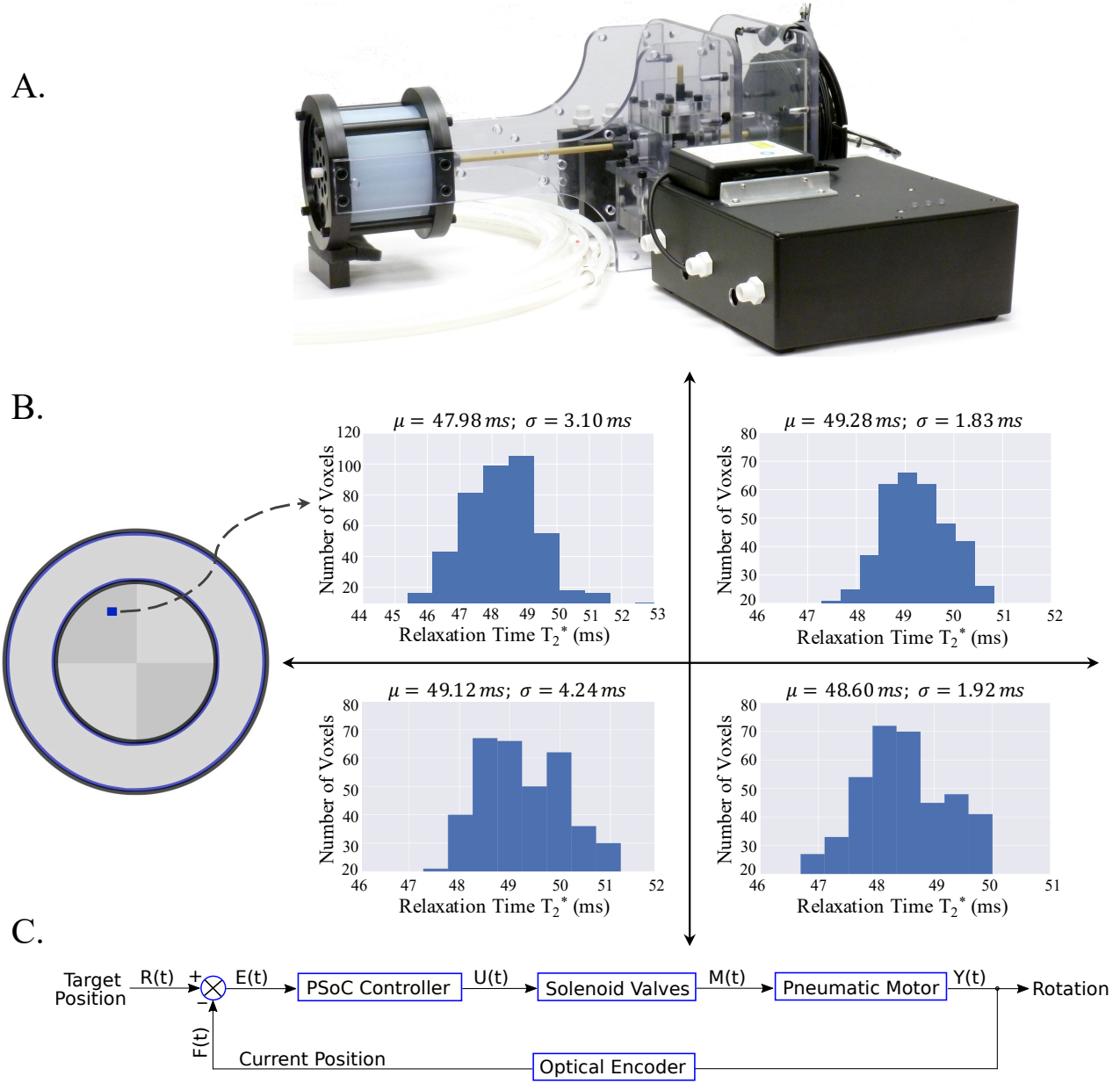

**Figure 1: A. Isometric view of the Dynamic Phantom**. The cylindrical head is the agarose gel cylinder assembly, which is coupled to a pneumatic motor and an optical encoder on the other end. All components remain intact via fastening to an outer frame and go inside the MR scanner with the cylindrical head placed inside the head-coil. The black box shown is the control unit, which interfaces with the optical encoder, pneumatic input from an air compressor and the pneumatic motor. **B. Distribution of T2* values across voxels in four quadrants at 3T (Site 1).** The agarose gel is prepared using the recipe provided by Friedman et al. (Friedman and Glover 2006). Even though the agarose gel is prepared only at 2.2% and 2.3% concentration, the heterogeneity in $T_2^*$ values can be attributed to imperfect agarose network formation, chemical heterogeneity, and polydispersity of gel networks(Djabourov et al. 1989). **C. Feedback control system for rotating the inner cylinder.** At each trigger from the MR scanner, the PSoC controller compares the current position F(t) with the programmed target position R(t) and opens the solenoid valve proportionally to the magnitude of the error signal E(t) to actuate the pneumatic

motor. Here, U(t) is the actuating signal, and M(t) is the manipulated variable. The system uses no braking mechanism, and accurate positioning is achieved through a predetermined linear relationship established between open-state time for solenoid valve and the corresponding rotation achieved at a given pneumatic pressure.

The phantom consists of three distinct parts: a) an agarose gel cylinder assembly, having two concentric cylinders; b) a control unit providing control logic for rotation of the inner cylinder; and, c) an air motor assembly with a gearbox and an optical encoder for position tracking. Within the agarose gel cylinder assembly, the inner cylinder rotates during the scan and is coupled to the air motor and the optical encoder, while the outer cylinder contains a reference gel and remains static. The outer cylinder's reference agarose gel is made at 2.2% concentration by weight, whereas the inner cylinder contains two different gel concentrations at 2.2% and 2.3% by weight, split into four quadrants in a configuration as shown in **Fig. 1B**. Within each quadrant, a variation in $T_2^*$ values exist across voxels because of imperfect agarose network formation, chemical heterogeneity, and polydispersity of gel networks (Djabourov et al. 1989). The control unit for driving the phantom uses a feedback control strategy with control logic implemented in PSoC microcontroller, feedback sensing via an optical encoder, and actuation through solenoid valves. The control unit contains some other custom circuitry for fast valve response time (spike-up voltage circuit), touchscreen user-interface running on raspberry-pi, and UART communication between the raspberry-pi and the PSoC microcontroller. The phantom is MR-compatible (agarose gel cylinder assembly and air motor assembly) and uses polycarbonate (body), delrin (air motor), glass-nylon (ball bearings), and G11 garolite (motor shaft) in construction. The control unit containing electronics and pneumatic compressor for driving the air motor stays outside in the MR control room.

**2.2.** *Using ground truth brain-like dynamic signals, we quantified a Standardized Signal-to-Noise Ratio (ST-SNR) and Dynamic Fidelity; these demonstrated wide variance across scanners, even for the "best case scenario" of high-performance scanners utilizing acquisition parameters individually optimized by a highly experienced MR physicist.* While the definition of signal-to-noise ratio (SNR) is well defined across the engineering domain, use of the term within the fMRI field has colloquially co-opted its definition in ways that can dilute its meaning and utility. Currently in fMRI, multiple definitions and variants for computing SNR exist (Welvaert and Rosseel 2013), leading to difficulty in interpreting and comparing SNR values. Normally used to optimize for scanner stability with the use of a static phantom, *temporal SNR* (tSNR) is defined as the ratio of mean signal to standard deviation of a time-series. However, for reasons described above, optimizing for tSNR (i.e., solely for stability) will suppress not only the fluctuations responsible for noise but also the fluctuations responsible for resting-state signal, effectively de-optimizing for detection of resting-state networks (DeDora et al. 2016). Furthermore, mean-signal in tSNR calculation is highly dependent on acquisition parameters, making the interpretation for comparison difficult. For example: tSNR has been reported across two orders of magnitude (e.g., between 4.42 and 280 for a recent review of studies (Welvaert and Rosseel 2013)). To address both issues, we quantified the accuracy with which fMRI time-series follow the true signal using two data-quality metrics: *Standardized Signal-to-Noise Ratio (ST-SNR)* and *Dynamic Fidelity*. "ST-SNR" is defined as the ratio of signal power and the background noise power and is calculated accordingly, where power is the sum of the absolute squares of time-domain samples divided by the time-series length. We define "*Dynamic Fidelity*" as the accuracy with which an MR scanner tracks changes in the input signal and calculate it as the Pearson correlation coefficient between the ground-truth signal and fMRI output.

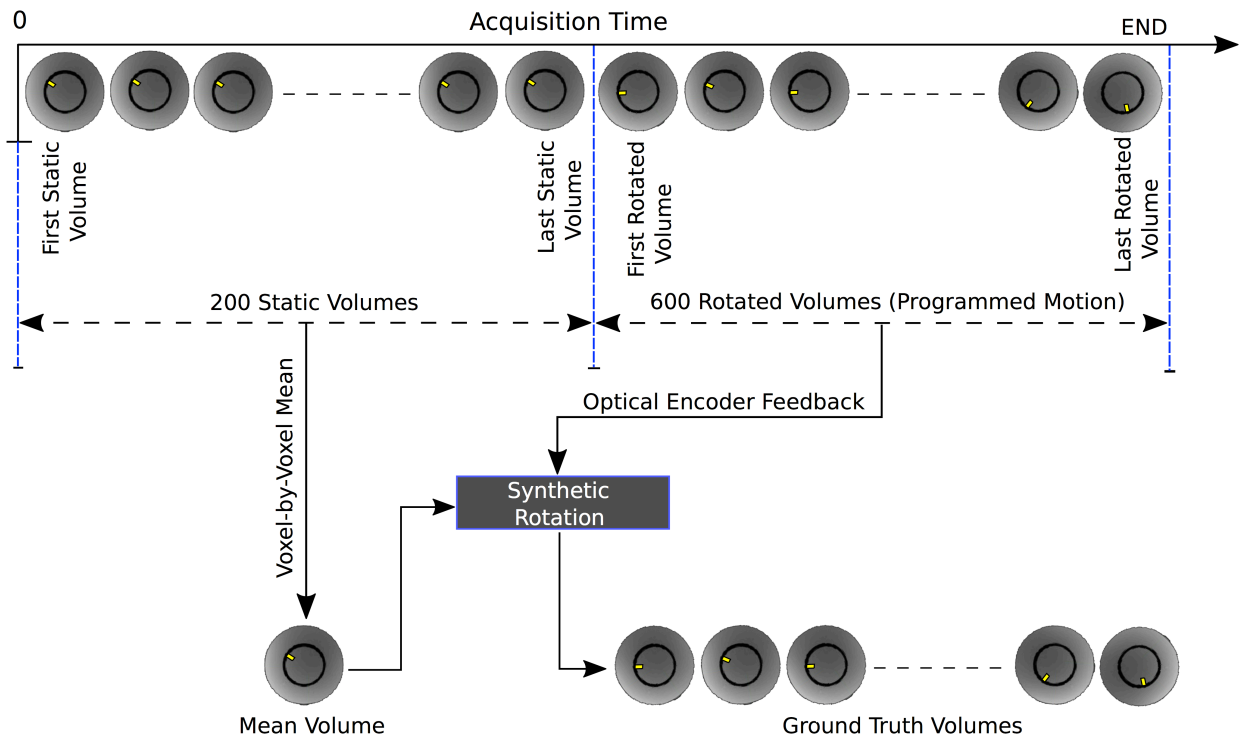

**Figure 2: Creating ground-truth using the Dynamic Phantom.** During each phantom scan, 200 static volumes were acquired and were averaged voxel-wise to obtain a close approximation to true intensity values. The mean volume was then rotated 600 times synthetically at angles obtained from the optical encoder during the actual run. This yielded ground-truth volumes, which then were compared to the volumes acquired during the scan.

The programmed rotation of the dynamic phantom, along with the optical encoder feedback, provides a mechanism for rotation control and sensing. The phantom tracks the programmed rotation at an accuracy of 0.2°. With the rotation generating voxel-wise time-series, the feedback sensing provides data on the actual rotation that occurs. This feedback data enables calculation of the ground-truth time-series and the noise estimate for each voxel, as shown in **Fig. 2**, for quantifying ST-SNR and Dynamic Fidelity. In **Table 1**, we show both ST-SNR and Dynamic Fidelity for four scanners, showing the potential for wide variance across scanners, even for a "best case scenario" of high-performance scanners utilizing acquisition parameters individually optimized by a highly experienced MR physicist. Importantly for multi-site or longitudinal applications, these two metrics (ST-SNR and Dynamic Fidelity) provide a direct assessment and comparison of data-quality over different scanners, as well as the same scanner over time. As ST-SNR and Dynamic Fidelity have standardized and interpretable range of values, the direct comparison of these metrics longitudinally or across scanners becomes possible. For example, for the same make and model of a scanner (the two Siemens PRISMA scanners, described in **Table 1**) having equivalent voxel-size, the ST-SNR observed is markedly different. Inspecting further, while one may attribute this difference to the different head-coil arrays used between the two scanners (**Table 3**), the comparison of ST-SNR with 3T SKYRA (**Table 1**) at the same site with equivalent voxel-size and head-coil suggests otherwise: that the Site 2 PRISMA scanner is an outlier.

| Scanner | Data Quality | | | Instability | Temporal Denoising | | | | | |
|---|---|---|---|---|---|---|---|---|---|---|
| | Fidelity | ST-SNR | Non-Linearity (% of Total Voxels) | | Contribution of Total Noise (%) | Bandpass Filter | | MP-PCA + Bandpass Filter | | CNN | |
| | | | Weak | Strong | | Fidelity | ST-SNR | Fidelity | ST-SNR | Fidelity | ST-SNR |
| Site 1: PRISMA 3T | 0.38 | 0.27 | 18 | 8 | 10.06 | 0.51 | 0.72 | 0.53 | 0.82 | 0.58 | 1.5 |
| Site 2: PRISMA 3T | 0.32 | 0.2 | 25 | 10 | 5.7 | 0.46 | 0.59 | 0.51 | 0.72 | 0.55 | 1.44 |
| Site 2: SKYRA 3T | 0.33 | 0.27 | 32 | 15 | 17.15 | 0.66 | 0.45 | 0.49 | 0.79 | 0.52 | 1.37 |
| Site 2: MAGNETOM 7T | 0.37 | 0.34 | 24 | 19 | 17.94 | 0.47 | 0.77 | 0.5 | 0.88 | 0.51 | 1.35 |

**Table 1:** Data-quality metrics for quality control, quantification of scanner-instability, and performance of bandpass filtering, MP-PCA and CNN temporal denoising scheme - evaluated for each scanner using phantom data. CNN denoising increases ST-SNR by ~4-7 times the measured time-series, bringing ST-SNR to a regime where signal power is higher than that of noise (ST-SNR > 1). Fidelity, ST-SNR and Instability are calculated on a single time series obtained after temporally concatenating voxels of interest.

**2.3. *Using the dynamic phantom generated ground-truth, we quantified the ratio of scanner instability to background noise in fMRI time-series, thereby identifying multiplicative versus thermal noise components.*** We analyzed time-series of the noise (residual time-series as calculated above, refer to **Fig. 2**), using power spectral density plots, to identify spectral-features arising from scanner artifacts. **Fig. 3** illustrates the mean power spectral density across all voxels for the ground-truth and the estimated noise time-series. Each voxel time-series' power spectral density was normalized by its maximum power before calculating mean at each frequency bin across all voxels. The power spectral density of the noise closely matches that of the ground-truth signal indicating the presence of multiplicative noise (scanner instability) component alongside thermal/background noise.

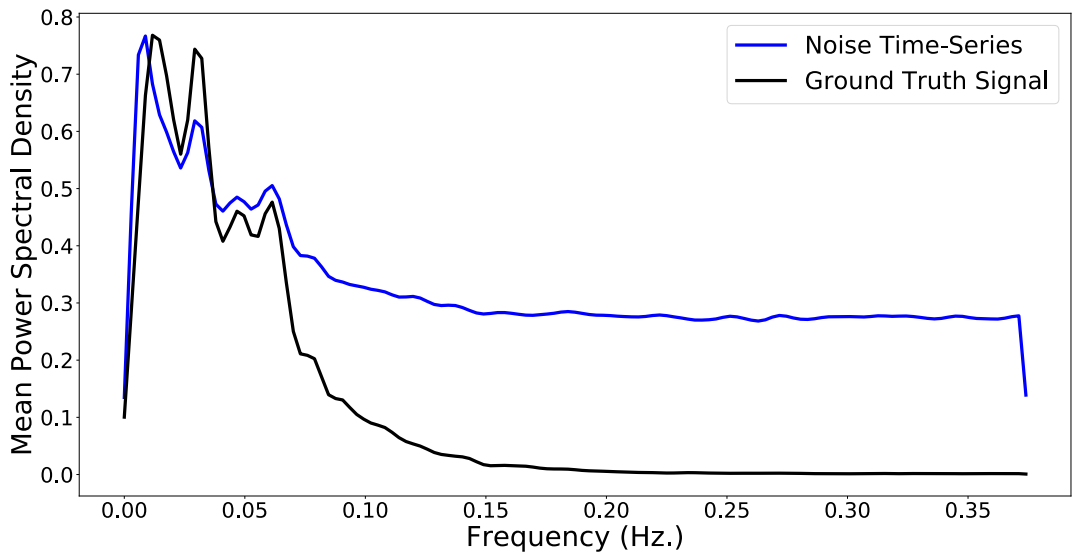

**Figure 3:** Qualitative comparison of mean power spectral density of the ground-truth and noise time-series provides signatures for the presence of signal-dependent (non-white) scanner confounds in fMRI data, in addition to the background noise. Voxel-wise noise time-series is calculated by subtracting measured fMRI time-series and the ground-truth time-series.

Multiplicative noise modulates the MR signal, is known to exhibit some temporal and spatial correlation (Greve et al. 2013), and cannot be removed using smoothing or frequency-based temporal filtering. The presence of multiplicative noise diminishes the advantages offered by hardware improvements (increase in signal to thermal noise ratio with higher field strength and more sensitive head-coil arrays) and can exacerbate the false-positives problem (Eklund, Nichols, and Knutsson 2016), alongside spurious correlations and poor reproducibility of functional connectivity. Band-limited programmed rotation of our phantom produces a band-limited ground-truth signal, and thus the associated multiplicative noise can be directly observed in this narrow band—see **Fig. 3** (in the 0–0.1 Hz range). At frequencies higher than 0.1 Hz where the ground-truth signal is absent, scanner noise shows a flat spectrum or white-noise behavior (thermal noise). This multiplicative noise behavior is further corroborated by a linear scaling of noise power (logarithmic scale) with an increase in signal (ground-truth) standard deviation. We observed a moderate correlation between noise power and signal standard deviation for all scanners (Site 1: PRISMA– r=0.35, Site 2: PRISMA– r=0.32, SKYRA– r=0.33, and MAGNETOM– r=0.35).

Using the ground-truth dynamic signal and the measured fMRI output, we quantified the ratio of multiplicative noise (scanner instability) to thermal/background noise using a probabilistic description of the two noise-sources. Scanner instability is signal-dependent and thus proportional to the signal intensity, while thermal noise is independent of the MR signal. Background noise and scanner-instability are temporally independent, and therefore, their variances add. With $\sigma_T$ as the standard deviation of the thermal noise and β as the proportionality constant for the multiplicative noise, we can write:

$$\sigma_{fMRI}^2 = \sigma_{GT}^2 + \sigma_T^2 + \beta^2 \sigma_{GT}^2 = \sigma_{GT}^2 + \sigma_{noise}^2, \qquad \text{Eq. 1}$$

where $\sigma_{fMRI}^2$ and $\sigma_{GT}^2$, are the variances of the observed fMRI output and the ground-truth, respectively. The model for the probability of observing the measured signal, given ground-truth $Y_{GT}$ and noise parameters can then be written as:

$$P(Y_{fMRI}|Y_{GT},\sigma_T,\beta) = N(\mu = Y_{GT}, \sigma^2 = \sigma_T^2 + \beta^2 Y_{GT}^2), \qquad \text{Eq. 2}$$

We estimate the parameters $\sigma_T$, and $\beta$ by Monte-Carlo simulation using $Y_{fMRI}$ and $Y_{GT}$. Specifically, we model **Eq.1** to sample from the posterior distribution that is proportional to **Eq.2** while assuming constant priors for the parameter distributions. The relative contribution of multiplicative noise to that of the total noise is listed in **Table 1** for each scanner. The results indicate that even in modern high-performance scanners with acquisition parameters optimized by a trained MR physicist, the scanner-induced variance due to instability is around 6–18% of the contribution of the total scanner noise. This range is consistent with Greve et al. (Greve et al. 2011), in which the authors measured scanner instability by scanning an agar phantom at two varying flip-angles to separate instability from background noise. Because we use different metrics, we included a detailed comparison between the Greve et al.'s (Greve et al. 2011) findings (supplementary materials, Table 1) and our study in the supplementary material. Finally, we provide a case-study comparing the two methods, using modern imaging hardware and acquisition parameters (multi-channel coils and parallel imaging) in the Supplementary Materials. We found agreement between the two methods, except when the background noise variance becomes space-variant. This suggests Greve et al.'s method risks inaccuracy for modern acquisition protocols, as previously discussed in Greve et al.

**2.4. *Using the dynamic phantom generated ground-truth, we quantified scanner-induced non-linearity in fMRI response.*** Finally, we observe scanner-induced temporal non-linear distortion of fMRI response using a tree-partition non-linearity estimator (Ljung 2019) (a piece-wise linear function defined by the binary tree over partitions of the regressor space) with ground-truth as the regressor. Non-linearity is detected in the observed fMRI data if a nonlinear function explains significant variance in the observed data beyond the variance explained by the linear function of the ground-truth. Non-linearity estimation was performed using 'isnlarx' function provided in System Identification Toolbox, Matlab (Ljung 2019), which categorizes non-linearity as strong, weak or not significant based on reliability of the nonlinearity detection test. We observed that the 7T scanner showed the highest non-linearity in response, with 19% of voxels exhibiting strong non-linearity (**Table 1**).

**2.5. *Using the dynamic phantom generated ground-truth, we evaluated the efficacy of applying random matrix theory to remove scanner-induced noise; thereby, demonstrating the utility of the dynamic phantom for comparing retrospective denoising techniques against a ground-truth.*** A method based on principal component analysis (PCA) coupled with random matrix theory (RMT), called MP-PCA(Veraart, Fieremans, and Novikov 2016; Veraart et al. 2016), has been introduced recently for denoising diffusion MRI(Veraart, Fieremans, and Novikov 2016; Veraart et al. 2016) and fMRI data (Adhikari et al. 2018). MP-PCA is a 4d image denoising technique that exploits redundancy in the PCA domain using the universal Marchenko–Pastur distribution to remove scanner-induced noise. MP-PCA denoising, followed by bandpass filtering in the frequency-band of interest (0.008-0.1Hz), showed increases in ST-SNR and Dynamic Fidelity over the observed fMRI data and the conventional bandpass filtering (0.008-0.1Hz). MP-PCA denoising showed a significant increase in Dynamic Fidelity with around 40%, 60%, 48%, and 35% increase and a ~2- to 3-fold increase in ST-SNR, for Site 1: PRISMA, Site 2: PRISMA, SKYRA, and MAGNETOM respectively, compared to the observed fMRI data.

**2.6. *We designed a data-driven temporal filter and observed robust increases in ST-SNR and Dynamic Fidelity of fMRI time-series after denoising.*** We provide a deep-learning framework using a Convolutional Neural Network (CNN) for learning an equivalent of a temporal filter. Given that we now have known dynamic inputs, we developed an end-to-end trainable CNN architecture that uses discriminative denoising to remove noise in the hidden layers. We provided pairs of measured fMRI time-series and known signal to learn a mapping from noisy to clean time-series implicitly. We used batch regularization with small batches of batch-size=8 within CNN to avoid internal covariate shift, accelerate the training process, and reduce dependence on network parameter initialization (Sergey Ioffe 2015). Sigmoid activation function has been used for non-linear mapping and a dropout layer for regularization (Nitish Srivastava 2014). The architecture details of CNN are specified in **Fig. 4**.

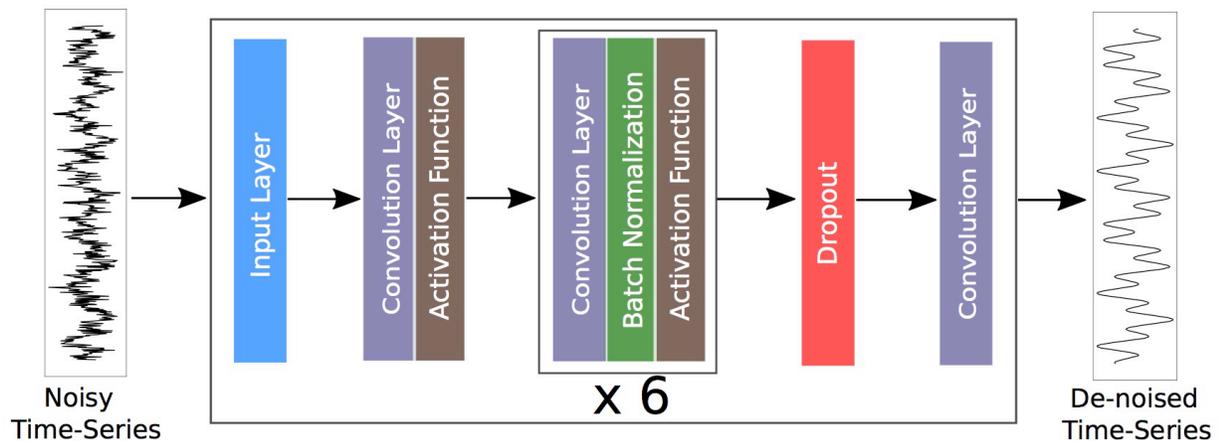

**Figure 4: Architecture of the convolutional neural network used for discriminative denoising.** Each convolution layer (except the last) contains 18 filters with a kernel size of 9 and a stride of 1. Sigmoid is used as the activation function. A dropout of 0.2 is used in the dropout layer. The last convolution layer contains only one filter. Negative of R-squared between the ground-truth and the denoised time-series used as the loss function (minimize) with Adam optimizer for stochastic optimization (Diederik P. Kingma and Ba 2014).

For evaluating the performance and generalizability of the CNN, we compare the results of CNN denoised fMRI time-series, as shown in **Fig. 5**, with the original data-quality and temporal de-noising using a standard third-order Butterworth bandpass filter (0.008–0.1 Hz). CNN de-noising showed a significant increase in Dynamic Fidelity with around 53%, 72%, 58%, and 38% increase, for Site 1: PRISMA, Site 2: PRISMA, SKYRA, and MAGNETOM respectively, compared to the observed fMRI data. Further, the CNN de-noising showed a ~4- to 7-fold increase in ST-SNR compared to the observed fMRI data. Finally, CNN de-noising outperforms the conventional temporal bandpass filtering and the MP-PCA denoising **(Table 1)** in terms of improving both the ST-SNR and the Dynamic Fidelity.

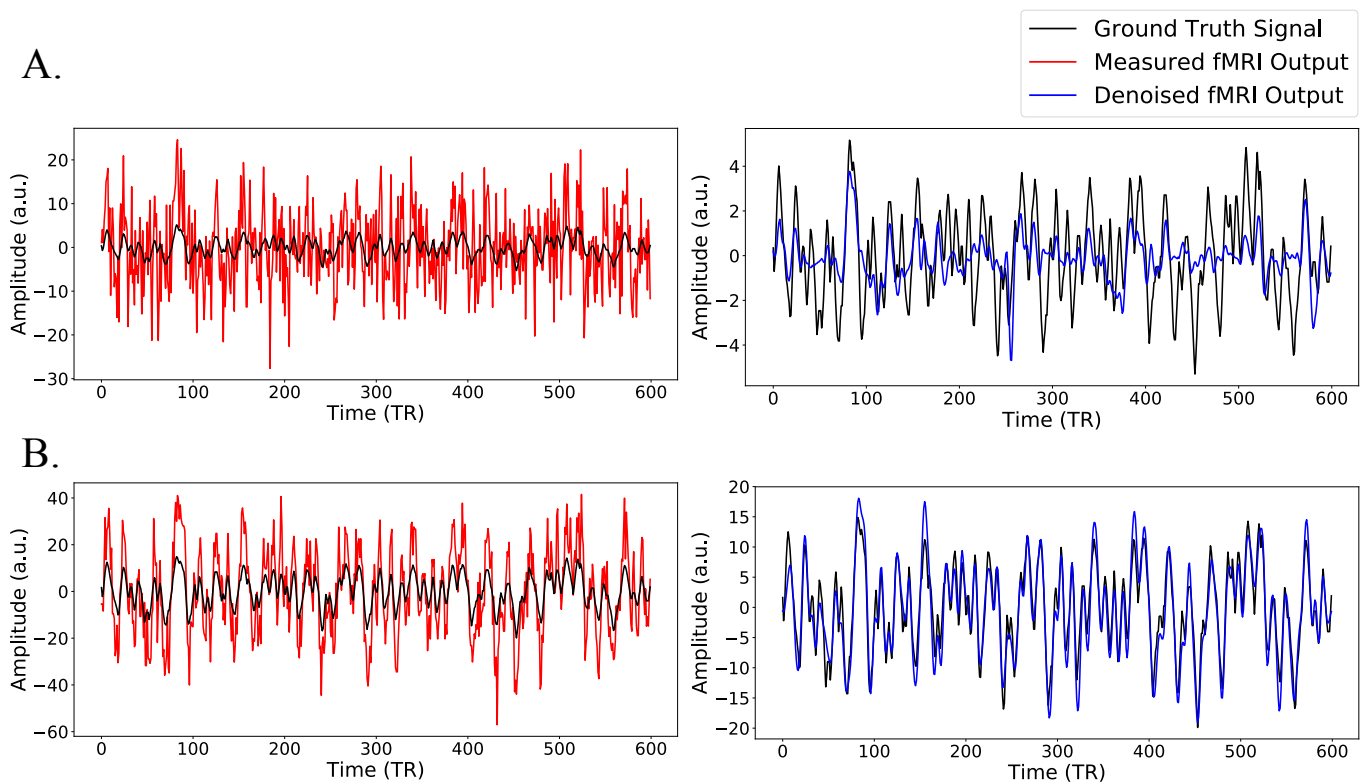

**Figure 5:** Exemplar denoising of fMRI output using the trained CNN for two voxels with **A.** low ST-SNR (0.06), and **B.** high ST-SNR (0.31) levels.

***2.7. Removing scanner-induced variance from human fMRI data increased the detection sensitivity of brain networks, visible even at the single-subject level.*** For assessing the effects of CNN de-noising on human fMRI data, the detection sensitivity of brain networks engaged in movie watching was calculated as a measure of the ability to preserve fluctuations of interest (signal) while removing scanner confounds (noise) from the time-series, and was quantified using the ratio of mean absolute Z-score inside and outside well-defined resting-state network masks in subject-specific ICA maps. A ratio > 1 indicates that Z-score inside the mask is higher compared to voxels outside. Higher this ratio, the easier it is to detect the brain/resting-state networks. We observed an increase in detection sensitivity at the single-subject level for all three scanners after accounting for scanner-related noise,

for both the MP-PCA denoising and the CNN denoising method. For MP-PCA denoising, permutation-testing revealed a significant increase in detection sensitivity for all three scanners (3T PRISMA: percent-change= 9.06% p-value=0.016; 3T SKYRA: percent-change= 13.03%, p-value=0.016; 7T MAGNETOM: percent-change=9.3%, p-value=0.015). Similar trends were observed for CNN denoising (3T PRISMA: percent-change= 13.63% p-value=0.016; 3T SKYRA: percent-change= 20.7%, p-value=0.015; 7T MAGNETOM: percent-change=18.74%, p-value=0.015). Furthermore, the CNN denoising outperformed MP-PCA denoising as evident from Table 2 (3T PRISMA: percent-difference = 4.19% p-value=0.016; 3T SKYRA: percent-difference = 6.78%, p-value=0.03; 7T MAGNETOM: percent-difference =8.64%, p-value=0.015).

|  | Subject # | 3T PRISMA ||| 3T SKYRA ||| 7T MAGNETOM |||
|---|---|---|---|---|---|---|---|---|---|---|
|  |  | Standard Method | MP-PCA Denoising | CNN Denoising | Standard Method | MP-PCA Denoising | CNN Denoising | Standard Method | MP-PCA Denoising | CNN Denoising |
| Run 1 | Subject 1 | 2.06 | 2.22 | **2.24** | 1.9 | 2.14 | **2.27** | 1.89 | 2.05 | **2.25** |
| Run 1 | Subject 2 | 2.25 | 2.45 | **2.55** | 2.11 | 2.39 | 2.39 | 2.37 | 2.61 | **2.85** |
| Run 1 | Subject 3 | 2.17 | 2.36 | **2.55** | 2.13 | 2.41 | **2.54** | 2.32 | 2.56 | **2.78** |
| Run 2 | Subject 1 | 1.88 | 2.01 | **2.04** | 1.95 | 2.25 | **2.38** | 2.12 | 2.24 | **2.44** |
| Run 2 | Subject 2 | 2.23 | 2.46 | **2.59** | 2.32 | 2.6 | **2.86** | 2.22 | 2.5 | **2.63** |
| Run 2 | Subject 3 | 2.1 | 2.34 | **2.45** | 2.25 | 2.52 | **2.84** | 2.42 | 2.62 | **2.89** |

**Table 2: Detection Sensitivity of resting-state networks.** Denoising of human fMRI data for removing scanner-confounds showed an increase in the detection sensitivity of resting-state networks. Here, detection sensitivity refers to the ratio of mean absolute Z-score inside and outside a well-defined resting-state network mask in subject-specific ICA maps. Detection sensitivity >1 indicates higher contrast of voxels-of-interest inside the brain network compared to voxels outside. The higher score is in **bold**.

## 3. Discussion

**3.1.** *Why should one use the dynamic phantom over a static phantom?* Static phantoms are commonly used for quality assurance (Friedman and Glover 2006) to assess and minimize scanner fluctuations due to background noise and instability. However, the resting-state fMRI or naturalistic paradigms depend not only upon suppressing fluctuations due to noise but equally upon sensitivity towards signal change, which can only be assessed by a phantom that produces a known and changing (dynamic) signal. The importance of a dynamic phantom is that it is the only method, to our knowledge, that can quantifiably assess the most basic assumption underlying all task-free fMRI:

fidelity between input (brain) dynamics and output (measured fMRI time-series) dynamics. We introduced a novel method for generating ground-truth using the dynamic phantom and estimating voxel-wise noise time-series. The dynamic phantom additionally provides an estimate of standardized signal-to-noise ratio (ST-SNR) and non-linearity, quantifying actual measurement error in fMRI response as compared to static-phantom derived temporal stability of the mean signal (tSNR). While static phantoms estimate only flat-spectrum noise (Expert et al. 2011), the dynamic phantom can detect both signal-dependent and background noise. Using Bayesian parameter estimation, we quantified the ratio of instability/multiplicative noise to the background noise. Although fMRI time-series have several sources of confounds and variance contributed by scanner-instability is relatively small, the reliability of the longitudinal data may be seriously affected without proper characterization. Using data metrics introduced, quality assurance protocols can be established for scanner health monitoring. Any deviations in ST-SNR, Dynamic Fidelity or scanner-instability, compared against longitudinally tracked measurements, would indicate scanner problems.

**3.2. *Why did we use a deep-learning approach for temporal denoising?*** Scanner-instability and background noise in resting-state data lead to decreased detection-sensitivity of resting-state networks, which have been typically addressed by increasing the amount of data collected or increasing the scan-time per subject. These methods are not only expensive but lead to other problems such as subject-fatigue and increased head-motion, which are especially acute in clinical populations. In the current report, we propose a fundamentally different approach for removing scanner confounds from fMRI time-series, which may circumvent the need for collecting more data, and which is ideally suited for single-subject level analyses required for clinical and computational modeling applications, as well as large-scale multi-site and longitudinal studies. Our method exploits the availability of paired measured fMRI and ground-truth data to perform discriminative denoising using CNN. Developing a denoising algorithm for correcting time-series distortions can be framed as a system-identification problem, wherein the goal is to infer a functional relationship between the system input (measured fMRI data) and the system output (denoised fMRI data). Convolution of the measured signal with the identified filter produces the denoised signal. While dealing with linear systems, this system-identification problem reduces to the characterization of impulse response using delta function or observing the system's frequency response using sinusoids. However, for non-linear systems, there exists no canonical representation of the system that will capture "all possibilities" of mapping inputs to transformed outputs. The convolution integral for linear systems can be extended to convolution-like Volterra series for non-linear systems, which can further be extended to Weiner series where each component of the series is orthogonal to all lower-order components. Lee and Schetzen (Lee and Schetzen 1965) provided a simple method based on cross-correlation for estimating Weiner kernels. However, the cross-correlation method is fundamentally limited by the fact that inputs must be Gaussian. Further, the kernel estimation suffers in cases of strongly nonlinear systems. To overcome these problems, we used deep learning for performing temporal filtering. Intuitively, the trained CNN can be thought of as a temporal filter (like a bandpass-filter), but with filter parameters estimated in an automated data-driven manner optimized for a specific scanner performing a session.

**3.3.** *Why is the dynamic phantom more useful than ICA-based techniques in mitigating scanner-effects for multi-site studies?* Different sites generally have very different-levels of scanner-noise (Greve et al. 2011), causing heteroscedasticity when using ordinary least-squares estimator and skewing the p-values to be smaller than they should be. Scanner-differences can be reduced by data-processing techniques before analysis (resting-state data), or scanner-effects can be adjusted statistically (task-based data). Feis et al.(Feis et al. 2015) recently showed the successful application of FMRIB's ICA-based X-noiseifier (FIX) (Salimi-Khorshidi et al. 2014) to remove scanner-specific structured noise components that diminished differences in detected resting-state networks across sites. However, the complexity in re-training the FIX classifier for a dataset from every new scanner is non-trivial and requires manual component labeling using data from multiple subjects by an expert. While our measurement of ST-SNR provides a way for statistical adjustment of scanner effects in task-based paradigm using ST-SNR as a covariate in ANCOVA designs, the CNN denoising can remove scanner-induced effects before analysis for resting-state fMRI or naturalistic paradigms in an automated way.

**3.4.** *Future directions:* Our work has direct implications in moving towards single-subject imaging, which is necessary for clinical purposes as well as for fMRI driven computational neuroscience. Ensuring the stability of time-series adds statistical power to draw useful conclusions from a limited amount of data. Although first-level analysis is dominated by physiological noise (Greve et al. 2011; Triantafyllou et al. 2005; Wald and Polimeni 2017), we observed a ~13–20% increase in detection sensitivity of resting-state networks after removal of scanner-related noise. The fact that the dynamic phantom can provide, for the first time, a ground truth, permits identification and removal of scanner-related noise. It also enables rigorous evaluation of new data-driven denoising methods under "real-world" conditions that may deviate from idealized *a priori* assumptions (i.e., physical models) of scanner noise characteristics. The dynamic phantom's optical encoder provides precise information (resolution = 0.04392 degrees) about phantom rotation, which can be used for evaluating both prospective and retrospective in-plane motion correction algorithms. Using the dynamic phantom for establishing data-quality metrics, will provide an evaluation of modern imaging protocols, for example compressed sensing fMRI or 3D EPI. Future studies with a larger sample-size will focus on the effects of removing scanner confounds on reliability estimates of functional connectivity analysis and computational neuroscience circuits. Low reliability causes low reproducibility of functional connectomics (Zuo, Biswal, and Poldrack 2019). Reproducibility across sessions while scanning the same patient affects the clinical decision making and thus is an active concern for the use of resting-state fMRI as a clinical tool (O'Connor and Zeffiro 2019). Further, as physiological noise, thermal noise, and scanner instability are temporally independent, the effect of physiological noise and scanner-induced fluctuations can be regressed out using a general linear model (GLM) framework. The second-order effects/interaction between physiological noise and scanner-induced fluctuations can easily be modeled using interaction terms in the GLM if external physiological recordings are available. The CNN output (denoised fMRI signal) and input (measured fMRI signal) can be used to obtain the regressors (subtracting denoised fMRI signal from the measured fMRI signal) for scanner-induced fluctuations, to model the interaction effects. Additionally, future directions include investigating effects of dynamic phantom estimated ST-SNR on activation effect size in task-based studies, combining multi-site task-based studies using ST-SNR as a covariate, and using CNN denoising to normalize data across sites as required for multi-site studies.

## 4. Methods

**4.1. *Study design:*** We performed imaging at two sites: the SCAN Center at Stony Brook University in Stony Brook, New York (Site 1) and the Athinoula A. Martinos Center for Biomedical Imaging at the Massachusetts General Hospital in Charlestown, Massachusetts (Site 2). We designed and engineered a dynamic phantom for producing ground-truth time series, based on differences in $T_2^*$ values of agarose gel across voxels of interest. Controlled rotation of the dynamic phantom produces variation in the $T_2^*$ values within a voxel, tuned to generate amplitude changes/signal as observed with BOLD contrast in humans (see Results for a detailed description of the design). At Site 1 (3T Siemens PRISMA scanner), we scanned the phantom during a single session with five acquisition runs, with each successive run separated by a 20-minute interval. Each run had a unique programmed rotation profile as input to the phantom. No human data acquisition occurred at Site 1. At Site 2, we acquired data from three human subjects (two males and one female aged 55, 56, and 47 years, respectively) and the phantom, using three scanners: 3T Siemens SKYRA, 3T Siemens PRISMA, and 7T Siemens MAGNETOM. We acquired data in three imaging sessions: one session per scanner. During each imaging session, we acquired three phantom scans, each with a unique rotation profile, and six human scans, with two scans per subject. The first phantom scan took place at the beginning of each session. Next, each of the three human subjects were scanned while they viewed a naturalistic movie (no audio, see Supplementary Materials for video) inside the scanner. Afterward, we acquired the second phantom scan, followed by a repeated acquisition for all three human subjects under identical conditions. Finally, we acquired the third phantom scan. The Institutional Review Board at Massachusetts General Hospital (Partner's Healthcare) provided approval for the human study, and all participants provided written informed consent prior to participating in the study.

**4.2.** *Data acquisition parameters:* To ensure that results conservatively reflect actual data-quality metrics within the neuroimaging field, we asked each scanner's MR physicist to independently provide the optimal acquisition parameters for modern fMRI studies conducted on that specific scanner. The details of the protocol parameters are as follows. (1) Site 1 (phantom imaging only): The phantom was scanned on a 3T Siemens PRISMA scanner with a 64-channel head coil. For relaxation rate measurements, multi-echo gradient-echo images were acquired at twelve echo times equally spaced between 5 ms and 60 ms with TR = 70 ms, FOV = 192 mm × 192 mm, flip angle = 20°, slice thickness= 1.5 mm, and readout bandwidth = 320 Hz/px. For the time-series data, standard single-shot gradient-echo EPI data were acquired with the parameters as listed in Table 3. (2) Site 2: Three different scanners were used for data acquisition. For phantom measurement, only EPI scans were acquired. For human measurements, structural scans based on a standard $T_1$-weighted MPRAGE and $B_0$ field maps were acquired in addition to the EPI scans. EPI scan parameters for all three scanners are listed in Table 3. Specifics of structural scans and $B_0$ field maps are: (a) 3T Siemens SKYRA: Structural scans, for spatial co-registration, were acquired as multi-echo MPRAGE with 1 mm isotropic voxel size and four echoes with $TE_1, TE_2, TE_3, TE_4$ = 1.69, 3.55, 5.41, 7.27 ms, TR= 2530 ms, flip angle = 7°, and GRAPPA acceleration =2. $B_0$ field map images, calculated using phase differences between gradient-echo images at TE = 3.47 ms and 5.93 ms, were acquired (TR = 500 ms, flip angle = 47°, voxel-size = 3.0 × 3.0 × 3.0 mm³ and 44 slices) for EPI distortion correction arising due to susceptibility-induced magnetic field inhomogeneity; (b) 3T Siemens PRISMA: Structural scans were acquired using a single-echo MPRAGE with 1 mm isotropic voxel size, TE= 2.9 ms, TR= 2500 ms, flip angle = 8° and GRAPPA acceleration= 2. $B_0$ field maps were acquired with TE= 3.47 and 5.93 ms, TR= 500 ms, flip-angle = 47°, voxel-size = 3 × 3 × 3 mm and 52 slices; (c) 7T Siemens MAGENETOM: Structural scans were acquired as multi-echo MPRAGE with 1 mm isotropic voxel size at four echoes with $TE_1, TE_2, TE_3, TE_4$ = 1.61, 3.47, 5.33, 7.19 ms, TR= 2530 ms, flip angle = 7°, and GRAPPA acceleration =2. $B_0$ field map images were acquired at TE = 4.60 and 5.62 ms, TR = 723 ms, flip angle = 47°, voxel-size = 1.7 × 1.7 × 1.5 mm³ and 89 slices.

| Parameter | Site 1 | | Site 2 | |
| --- | --- | --- | --- | --- |
| Scanner | Siemens PRISMA | Siemens PRISMA | Siemens SKYRA | Siemens MAGNETOM |
| B0 Field | 3T | 3T | 3T | 7T |
| Head Coil | 64 | 32 | 32 | 32 |
| TR (msec) | 1000 | 800 | 748 | 802 |
| TE (msec) | 33 | 30 | 31 | 20 |
| Flip Angle (degrees) | 52 | 52 | 52 | 33 |
| EPI Factor | 84 | 90 | 80 | 96 |
| Voxel Size | 2.5mm Isotropic | 2.4mm Isotropic | 2.5mm Isotropic | 2 mm x 2mm x 1.5mm |
| Number of Slices | 28 | 60 | 48 | 85 |
| Number of Volumes* | 800/600 | 800/600 | 800/600 | 800/600 |
| Echo-Spacing | 0.58 | 0.51 | 0.59 | 0.55 |
| iPAT | 1 | 1 | 1 | 2 |
| Multiband Factor | 4 | 6 | 6 | 5 |
| Bandwidth (Hz/Px) | 2990 | 2778 | 2232 | 2368 |

**Table 3:** Acquisition parameters for functional EPI datasets for both the phantom and human subjects. *Only 600 volumes acquired in the case of human subjects.

**4.3. *Preprocessing of phantom data for calculating data quality metrics and training the convolutional neural network (CNN):*** Acquisition of phantom EPI data involved acquiring the first 200 volumes without any programmed rotation, followed by 600 rotating volumes with the rotation synchronized to the scanner's TR (repetition time) trigger signal. The phantom rotation was limited to around 250 ms from the start of each TR and was quantified using the optical encoder's feedback (Figure 1A, C). Before analysis, we corrected all phantom acquisitions for smooth spatial intensity variations caused by nonuniformity in the B0 field, $B1^+$ field, and receiver coil sensitivity (Sled and Pike 1998; Sled, Zijdenbos, and Evans 1998) using the N4ITK algorithm (Tustison et al. 2010), implemented in ANTs toolbox. N4ITK offers improved bias field correction over the original nonuniform intensity normalization (N3) algorithm (Sled, Zijdenbos, and Evans 1998), via robust b-spline approximation and a hierarchical optimizer to model a range of bias modulation. Based on the optical encoder's feedback and scanner's slice timing information, all the slices acquired during in-plane rotation within a TR were discarded from the respective EPI dataset for any further analysis. The remaining slices were manually inspected, and bad slices due to susceptibility artifacts (towards the top and bottom face of the cylinder) were thrown out. The final set of slices then underwent an automated procedure based on contour finding and the Hough transform for generating masks used to select the voxels of interest located in the inner cylinder of each slice. The first 200 volumes of all the remaining slices were averaged voxel-wise to create a mean functional dataset to obtain close approximations to the true voxel intensity. Synthetic rotation of the mean functional dataset, to create ground-truth time-series, involved up-sampling the mean images by a factor of 5 (3rd order spline interpolation), followed by rotation at angles provided by optical encoder's feedback and down-sampling by local averaging to original dimensions of the mean functional slice. Subtracting the noisy fMRI output from the corresponding ground-truth time-series yields voxel-wise noise time-series. Power spectrum density (Fig. 3) was calculated using the welch method implemented in SciPy library (Virtanen et al. 2020). Monte-Carlo simulations for parameter estimation to quantify multiplicative-to-thermal noise ratio were carried out in PyMC3 (Salvatier, Wiecki, and Fonnesbeck 2016). We estimated the percentage of voxels exhibiting nonlinearity for each scanner. For a given voxel with ground-truth time series G(t) and a measured fMRI time series Y(t), we express the measured fMRI time series as

$$Y(t) = L(t) + F(t) + E(t)$$

where L(t) represents the portion of data explained by a linear function of the ground-truth time series, F(t) represents the portion of data explained by a nonlinear function of the ground-truth time series and E(t) represents unexplained residual variance. If the nonlinear function explains a significant portion of variance after regressing out the linear model L(t) from Y(t), a nonlinearity is detected in the time series Y(t) (Ljung et al. 2006; Sjöberg et al. 1995). F(t) models the nonlinearity based on a nonlinear function/estimator (Sjöberg et al. 1995; Ljung et al. 2006) of the ground-truth time series, which can be a binary partition tree, a radial basis function network based on wavelets, a piecewise linear estimator, a multi-layer neural network or custom-built non-linearity regressors (for example, quadratic or polynomial regressors of ground-truth time series). We used a binary tree partition (Vanli and Kozat 2014; Ljung et al. 2006) as nonlinearity estimator, which splits the data into two subsets followed by iterative splitting of each subset into smaller subsets to partition the entire regressor space (ground-truth time series) into a binary tree. After this, linear regression is performed at each level of the binary tree to complete the estimation procedure (Vanli and Kozat 2014). We performed the nonlinearity estimation with a binary partition tree using the "isnlarx" function provided in System Identification Toolbox, Matlab (Ljung 2019).

For all voxels, the measured and the ground-truth time-series pairs were used for end-to-end training of the CNN (see Fig. 4 for architecture). Given that multiple phantom scans were acquired for each scanner, CNN training involved combining data acquired with different programmed motion sequences (Supplementary Materials: Suppl. Fig. 2) on a scanner for data-augmentation. Within each training dataset, 33% of data was used as validation split and model weights with lowest validation loss was saved as the trained CNN. For Site 1, three CNNs were trained using data from scans 1 and 3, scans 2 and 4, and scans 3 and 5. For Site 2, three phantom scans were acquired at each scanner, and CNNs were trained using data from scans 1 and 2, scans 2 and 3, and scans 1 and 3. For testing denoising performance, the test data were denoised using a trained CNN which did not use the test data during training (out-of-sample denoising), for example: at Site 2, for denoising scan 2, we used a CNN trained on scans 1 and 3.

**4.4. *Preprocessing of Human Data:*** Spatial preprocessing was performed in the Statistical Parametric Mapping (SPM12) software package (http://www.fil.ion.ucl.ac.uk/spm) using the pipeline provided in the CONN toolbox (Whitfield-Gabrieli and Nieto-Castanon 2012). Functional images were motion (rigid alignment, six-degrees-of-freedom) and $B_0$ field map corrected, and a mean functional image was calculated for each subject. The mean functional images were then co-registered to high-resolution structural images followed by segmentation to generate gray matter, white matter, and cerebrospinal fluid images. Each voxel time-series was demeaned and underwent quadratic de-trending. For further temporal preprocessing, the data went through two different pipelines to generate three datasets as discussed below: (a) <u>Standard Method:</u> Physiological confounds were removed using the Component-Based Noise Correction Method (Behzadi et al. 2007) (CompCor) implemented through Nipype interface (Gorgolewski et al. 2011). CompCor regresses out the confounding effects of multiple empirically estimated noise sources calculated from variability in BOLD time-series of cerebrospinal fluid and white matter (based on principal component analysis). Five components of white matter and cerebrospinal fluid, and six motion parameters, along with temporal bandpass filtering (0.008–0.1 Hz), were used for physiological denoising. Removal of confounds was orthogonal to the bandpass filtering(Lindquist et al. 2019); (b) CNN Denoising: Spatially preprocessed functional data (motion and fieldmap corrected and normalized to MNI) underwent denoising (voxels in gray-matter only) using trained scanner-specific CNN, followed by physiological confound removal as in the standard method (CompCor, motion, and bandpass filtering); (c) MP-PCA denoising: We repeated the spatial preprocessing of functional data and applied the standard method of temporal preprocessing, on MP-PCA denoised raw functional data, to generate a third dataset in addition to the standard method and CNN denoising datasets. Finally, datasets obtained from all three denoising methods were smoothed with a 4-mm full width at half-maximum Gaussian kernel, followed by normalization to $2 \times 2 \times 2$ mm Montreal Neurological Institute (MNI) EPI template.

**4.5.** *Calculating detection sensitivity of resting-state networks:* To identify functionally connected networks in a data-driven manner, we performed group spatial ICA on the preprocessed data using the GIFT v3.0b fMRI Toolbox (https://trendscenter.org/software/), separately for each scanner and temporal processing scheme (standard method and CNN denoising). For each dataset, 20 independent components were obtained, after ten runs of ICASSO (Himberg, Hyvärinen, and Esposito 2004) procedure for ensuring component stability. Subject-specific spatial maps and associated time courses were estimated using back-reconstruction (GICA) (Erhardt et al. 2011). We used the Infomax algorithm for performing ICA. ICA spatial maps were converted to Z values. We spatially matched the subject-specific ICA maps to seventeen well-defined resting-state network templates obtained from Yeo et al. (Yeo et al. 2011), for obtaining each subject's corresponding network ICA maps. Detection sensitivity was then calculated as the ratio of mean absolute Z-score inside and outside of each of the seventeen resting-state network masks applied to the matched subject-specific ICA spatial maps. The mean of detection sensitivity values, across all seventeen networks, for each subject, yielded a total of six values (three subjects with two runs) for every scanner. These six values were compared between the standard method, MP-PCA denoising, and the CNN temporal denoising for each scanner using permutation testing (100,000 repetitions).

## Declaration of Interests

R.K, A.L, J.R.P declare no conflict of interest. The intellectual property covering the dynamic phantom is owned by the Research Foundation for The State University of New York (RF SUNY). L.R.M.-P and H.H.S, as inventors, will receive a share of the royalties under the terms prescribed by the RF SUNY. ALA Scientific Inc., NY – USA, is the licensee for commercialization of the dynamic phantom. L.T declares employment, and A.K declares employment & ownership in ALA Scientific Inc.

## Acknowledgements

The research described in this paper was funded by the National Science Foundation (STTR Phase 1 Grant 1622525), the National Institute of Drug Abuse (SBIR Phase 1 & Phase 2 Grant 1R44 DA043277-01) and the National Heart, Lung and Blood Institute (Grant 5U01HL12752202). The authors gratefully acknowledge the assistance provided by Jie Ma and Michael Bishop during design and engineering of the dynamic phantom, and Anar Amgalan for discussions on methodology and software development.

## CRediT Authorship Contribution Statement

**Rajat Kumar:** Conceptualization, Methodology, Software, Formal Analysis, Investigation, Resources, Writing - original draft, Writing - review & editing. **Liang Tan:** Software, Resources. **Alan Kriegstein:** Resources, Funding acquisition. **Andrew Lithen:** Investigation. **Jonathan R. Polimeni:** Supervision, Writing – review & editing. **Helmut H. Strey:** Conceptualization, Methodology, Software, Resources, Supervision, Writing – review & editing, Funding acquisition. **Lilianne R. Mujica-Parodi:** Conceptualization, Methodology, Supervision, Writing – original draft, Writing – review & editing, Project administration, Funding acquisition.

## Data and Code Availability Statement

The data used in this study are available from OpenNeuro upon request to the corresponding authors, without requiring any formal data-sharing agreement. For data analysis, SPM can be obtained from - https://www.fil.ion.ucl.ac.uk/spm/; Conn-Toolbox can be obtained from - https://web.conn-

toolbox.org/; GIFT can be obtained from - https://trendscenter.org/software/; custom-software code used in the analysis can be obtained from - https://github.com/RajatKGupta/fMRI_BrainDancer.

# Supplementary Material

## 1. Scanner instability comparison with Greve et al.'s study (Greve et al. 2011):

Greve et al. did not report exact instability numbers for phantom measurements, and therefore we used the plots provided in Greve et al. (figure 1B and 2B), to perform the calculations. As per the definitions in Greve et. al, signal-weighted signal-to-fluctuation-noise (swSFNR) and background signal-to-fluctuation-noise (bgSFNR) can be written as:

$$swSFNR = \frac{\mu_\alpha}{\sigma_{SW_\alpha}} \quad (1)$$

$$bgSFNR = \frac{\mu_\alpha}{\sigma_{bg}} \quad (2)$$

where $\mu_\alpha$ is the intensity of a voxel averaged over time, $\sigma_{SW_\alpha}$ is standard deviation of signal-dependent multiplicative noise at flip angle $\alpha$ and, $\sigma_{bg}$ is background/thermal noise standard deviation.

Using swSFNR and bgSFNR from figure 1B and 2B in Greve et al., we can calculate the percentage contribution of multiplicative noise using equations (3) and (4):

As, $$\frac{bgSFNR^2}{swSFNR^2} = \frac{\sigma_{SW_\alpha}^2}{\sigma_{bg}^2} \quad (3)$$

Therefore,

$$\% \text{ Mutliplicative Noise or } \% \text{ Scanner Instability} = \frac{\sigma_{SW_\alpha}^2}{\sigma_{bg}^2 + \sigma_{SW_\alpha}^2} * 100 \quad (4)$$

The estimates of scanner instability obtained for phantom measurement from Greve et al. are reported in Table 1 and are comparable to reported measure of instability (5.7% - 17.94%) in our study.

| Site | Scanner Instability |
|---|---|
| Site 03 | 20% |
| Site 05 | 8.25% |
| Site 06 | 2.91% |
| Site 18 | 2.34% |

**Table 1:** Percentage contribution of scanner instability of total noise for Greve et al. (Greve et al. 2011), computed on phantom measurements.

## 2. Case-study comparing Greve et al.'s method with the probabilistic model (section 2.3. – main text) for calculating scanner instability.

*Materials and Methods:* We provide a direct comparison between Greve et al.'s method (Greve et al. 2011) and our probabilistic model for calculating scanner instability, for the 3T SKYRA and the 7T MAGENTOM at their operating flip-angles. We used the fBIRN agar phantom (Greve et al. 2011; Friedman and Glover 2006) and acquired datasets at two different flip angles for implementing Greve et al.'s method. We performed the first acquisition at a flip angle of 10 degrees (same as Greve et al.). We performed the second acquisition at a flip angle of 52 degrees for the 3T SKYRA and 33 degrees for the 7T MAGNETOM, to match the acquisition protocol used in our study. As signal-weighted multiplicative noise scales from flip angle $\alpha_1$ to $\alpha_2$, Greve et al.'s method utilizes the equations (1-3),

$$M = \frac{\mu_{\alpha 1}}{\mu_{\alpha 2}} \; ; \; \sigma^2_{SW,\alpha 2} = \frac{\sigma^2_{SW,\alpha 1}}{M^2} \quad (1)$$

$$\sigma^2_{SW,\alpha 1} = \frac{M^2(\sigma^2_{\alpha 1} - \sigma^2_{\alpha 2})}{M^2 - 1} \quad (2)$$

$$\sigma^2_{bg} = \frac{M^2 \sigma^2_{\alpha 2} - \sigma^2_{\alpha 1}}{M^2 - 1} \quad (3)$$

to estimate signal-weighted multiplicative noise at $\alpha_2$, where $\mu_{\alpha i}$ is the mean intensity, $\sigma^2_{\alpha i}$ is the total variance, $\sigma^2_{SW,\alpha i}$ is the signal weighted variance at the flip angle $\alpha_i$ and $\sigma^2_{bg}$ is the background noise. We then acquired data at the second/operating flip angle using our dynamic phantom for calculating scanner instability. All data was acquired in a single session. For analysis, we used the methods as in Greve et al. In brief, we created a 12 cm (diameter) spherical 3D ROI and extracted voxel-wise time series within the ROI. For each voxel, we fit a second-order polynomial; the constant term in the polynomial was used as an estimate of the mean, and the residual variance was used as an estimate of noise variance at the voxel. The mean and variance maps of the ROI were then used to compute the mean and noise variance across all voxels (average) for each flip angle, followed by estimation of instability at the operating flip angle. For comparison, we estimated instability using the dynamic phantom as per our probabilistic model.

*Results and Discussion:*

a) Greve et al.'s method and our probabilistic model produced comparable instability estimates for the 3T scanner (iPAT factor =1; multi-band acquisition with 32 channel head coil). Estimated instability was 16.40% and 15.82% of the total noise variance using the Greve et al.'s method and our probabilistic model, respectively.

b) Greve et al.'s method and our probabilistic model produced comparable instability estimates for the 7T scanner (iPAT factor= 2, 32 channel head coil), only when small ROIs were used for instability calculation in the Greve et al.'s method. With larger ROIs, Greve et al.'s method incorrectly estimates instability. Estimated instability for a 2 cm ROI was 32.48% of the total noise variance using the Greve et al.'s method and was comparable to 33.01% instability estimate using our probabilistic model. Table 2 shows the variation in estimated scanner instability using Greve et al.'s method for varying ROI diameters.

| ROI Diameter | Instability |
|---|---|
| 2 cm | 32.48% |
| 4 cm | 47.99% |
| 12 cm | 62.59% |

**Table 2:** Variation in estimated scanner instability using Greve et al.'s method for varying ROI diameters.

Greve et al. very clearly state in the discussion section of their manuscript (Greve et al. 2011) that – *"One limitation in this method is that if the noise variance changes over the ROI, then the method may yield inaccurate results…..Parallel imaging changes the demands made on the scanner hardware, and parallel imaging reconstruction methods will likely impart spatial variation in the noise variance. In these cases, the use of smaller ROIs may be required"*.

Because of this change in variance with ROI size, Greve et al.'s method is problematic to implement with parallel imaging due to the absence of an accurate background noise estimate. Our method avoids problems arising due to the spatial variation of the noise variance because we measure the ground-truth signal for each voxel of the inner-cylinder of the phantom. Noise estimation is thus based on a direct comparison between ground-truth versus measured signal for each voxel.

**Table 3: Bayesian parameter estimates for instability-to-background noise ratio calculation.** Posterior distributions for β (proportionality constant for the multiplicative noise) and $\sigma_T$ (standard deviation of the thermal/background noise) normalized to ground truth, obtained from Monte-Carlo simulation.

| Scanner | Parameter Estimates | |
|---|---|---|
| | $\beta$ | $\sigma_T/\sigma_{GNT}$ |
| Site 1: PRISMA 3T | 0.613 ± 0.002 | 1.83 ± 0.001 |
| Site 2: PRISMA 3T | 0.541 ± 0.003 | 2.20 ± 0.001 |
| Site 2: SKYRA 3T | 0.804 ± 0.003 | 1.77 ± 0.001 |
| Site 2: MAGNETOM 7T | 0.730 ± 0.001 | 1.56 ± 0.001 |

**Figure 1**: *(Left)* **Dimensions of the inner-cylinder and** *(Right)* **the outer-cylinder (in [mm] and inches).** The inner cylinder weighs 217g in total, with 138g of agarose gel. The outer cylinder weighs 807g in total, with 580g of agarose gel.

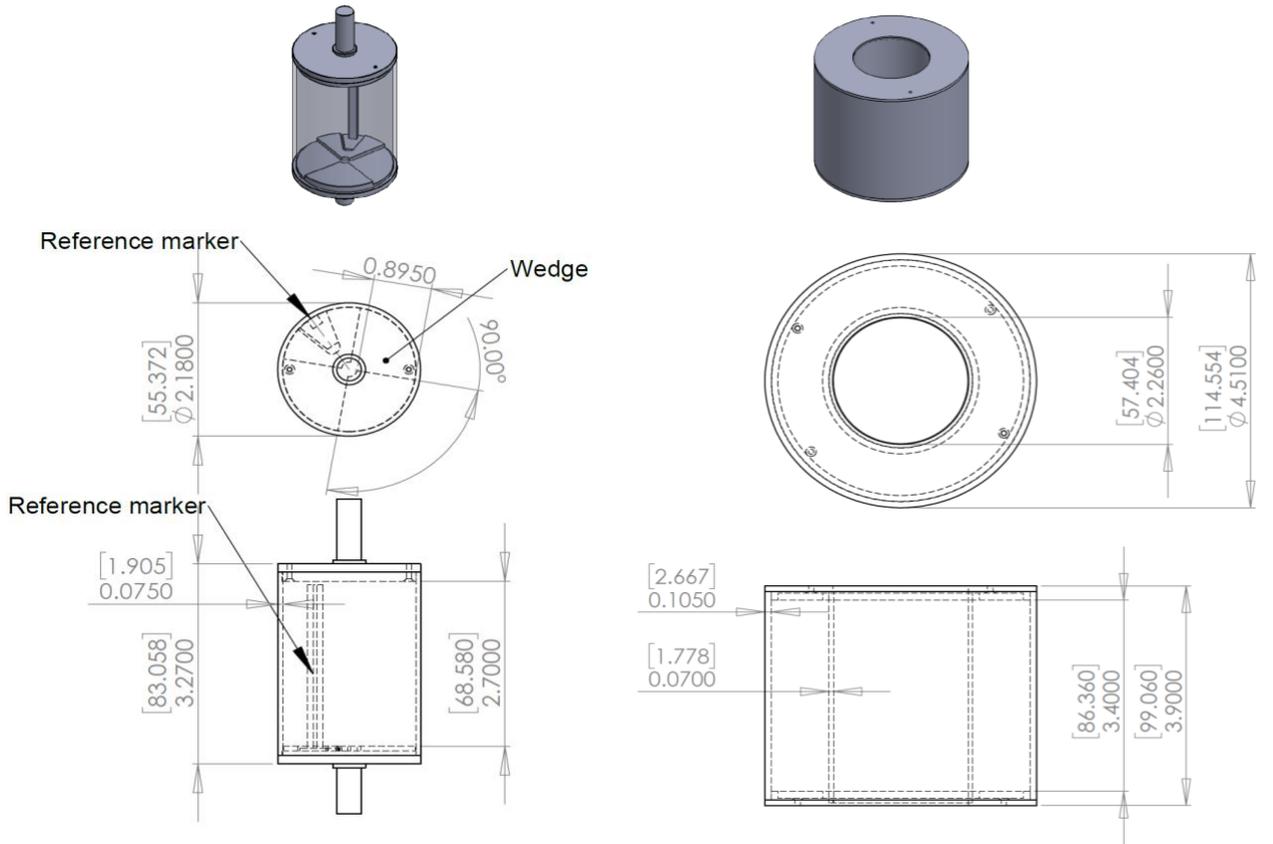

**Figure 2: Power spectral density of five brain-like dynamic signals programmed in PSoC microcontroller for phantom rotation**. Variations in programmed rotation sequences were used for data augmentation while training the CNN, as well as testing generalizability of the trained CNN for a given scanner. Sequence 1-3 were used at both sites, while sequence 4,5 were used only at site 1.

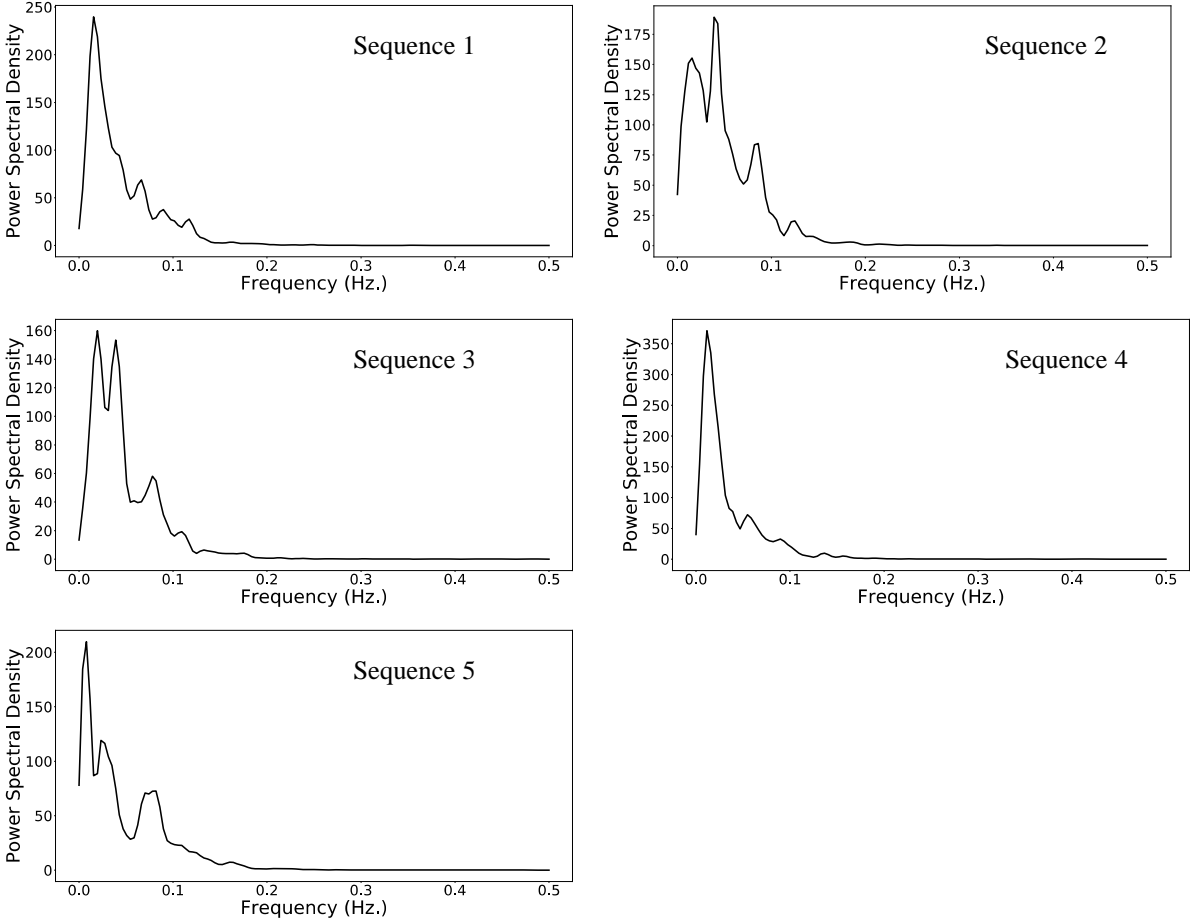

**Figure 3: Distribution of detection sensitivity values for 17 well defined resting-state networks across various denoising methods.**

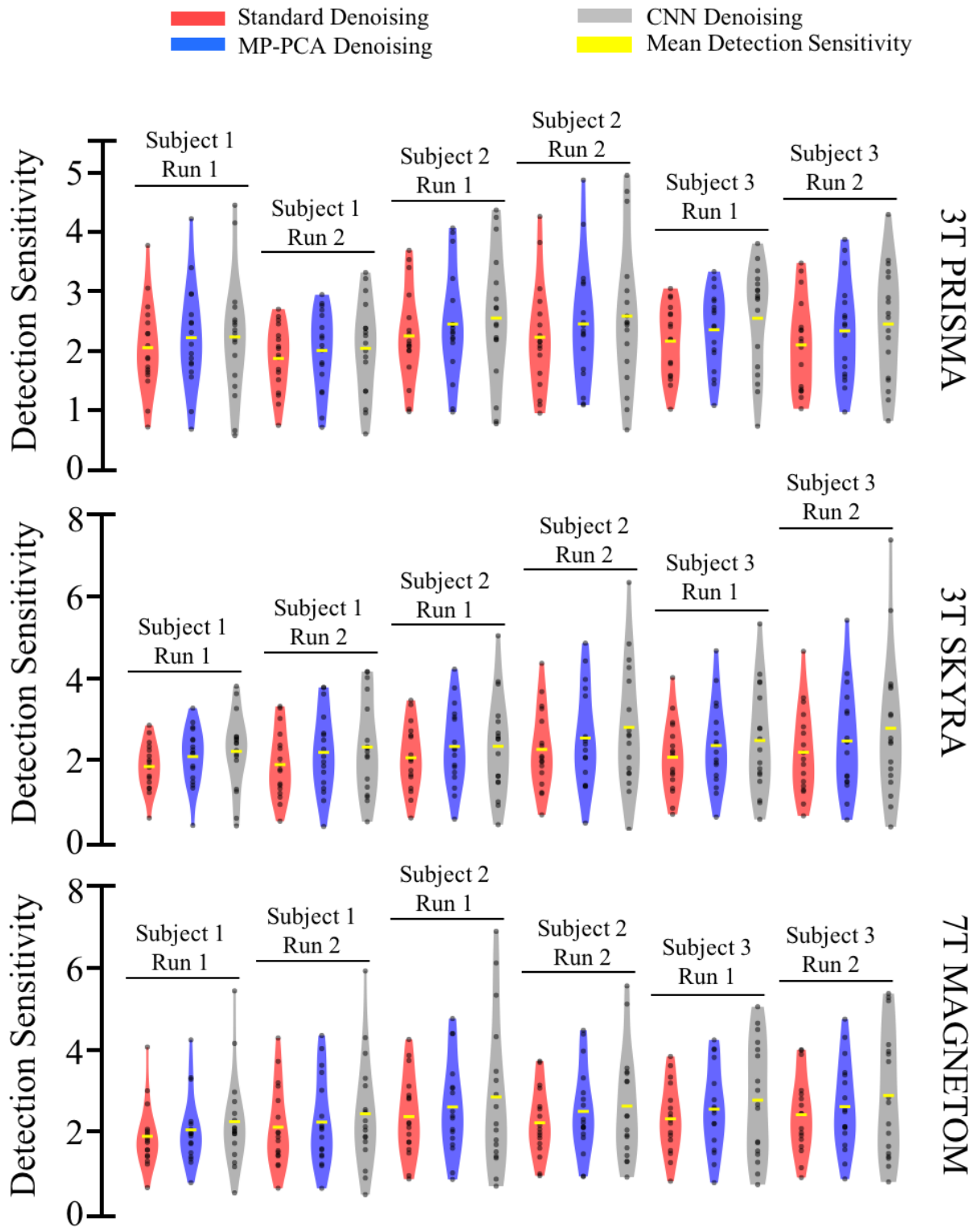

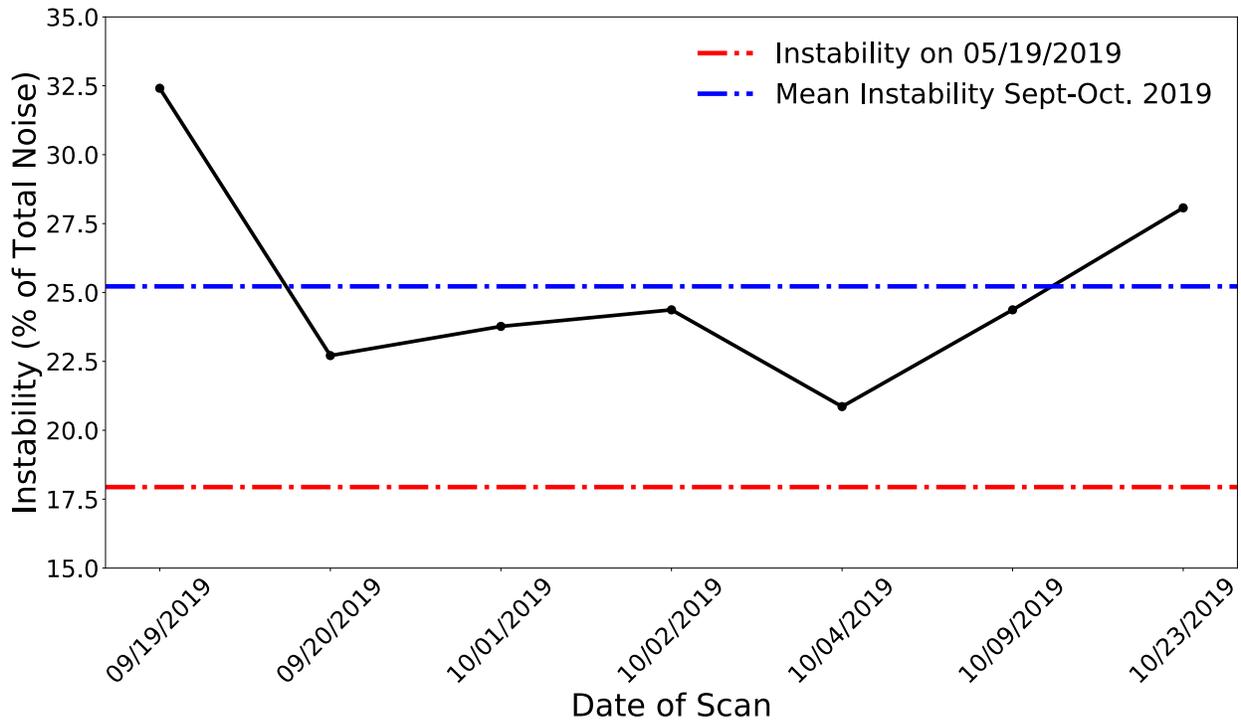

**Figure 4: Scanner Instability for 7T Magnetom.** The data used in the current manuscript was acquired on 05/19/2019. We acquired some QA scans during September-October 2019 on the same scanner with same acquisition parameters. There is a clear indication of scanner issues/increasing scanner instability between May and September 2019. Additionally, variance in scanner instability can be noticed on a daily basis.

Figure 5 – 10: ICA decomposition (20 components) for all three scanners at Site 2. Download [here](here).